# Grain boundary and defects assisted thermal conductivity of nano-crystalline $Gd_2Ti_2O_7$ Pyrochlore

Yogendar Singh[1], Vivek Kumar[1], Saurabh Kumar Sharma[1], P. K. Kulriya[1,a]

[1]School of Physical Sciences, Jawaharlal Nehru University, New Delhi 110067, India

**Abstract**

The thermal conductivity study on the pyrochlore structured ceramics is important for utilization of these materials as an inert matrix fuel, electrolytes for oxide fuel cell and thermal barrier coating. The impact of porosity, structural defects and boundary scattering on thermal properties of the nanocrystalline $Gd_2Ti_2O_7$ fabricated by spark plasma sintering followed by high energy ball milling has been investigated. The thermal conductivity has been measured in the temperature range from room temperature to 900 °C using laser flash apparatus (LFA), and results were discussed by including several phenomena contributing to the thermal transmission in nc-pyrochlore in support of the scattering mechanism. A systematic decrease in thermal conductivity with reducing grain size can be understood by the phonon-pore scattering phenomenon. Furthermore, dominate of the boundary scattering on the thermal transmission with the increasing temperature is also observed. The experimentally determined values of thermal conductivity are compatible with the thermal insulation requirements for thermal barrier coating (TBC) applications.

**Highlights:**

[a]Author to whom correspondence should be addressed.

Present address: School of Physical Sciences, Jawaharlal Nehru University, New Delhi 110067, India, electronic mail: pkkulriya@mail.jnu.ac.in

**Introduction**

Thermal conductivity of pyrochlore structured oxides with composition $A_2B_2O_7$ is of quite interest due to their utilization as inert matrix material to simulate the transmutation of minor actinides in the advanced nuclear reactor [1,2], thermal barrier coatings (TBCs) [3], electrolytes for solid oxide fuel cells [4,5], sensor applications [6] and immobilization of high-level nuclear waste [1,7]. In particular, for thermal barrier coatings and inert matrix importance, ceramics must have unique physico-chemical properties relevant to behavior under extreme conditions. Generally, thermal barrier coatings are proposed to blades and other components of hot environment in jet engines to protect the base materials (Si-based ceramics, super-alloys) against harsh environment and high temperatures to rise the operating temperature of gas turbine engines [8]. The vital basic requirements of materials for TBCs are very high melting point, low thermal conductivity, resistance to sintering at high temperatures, and phase stability at operating temperature. The low thermal conductivity is a critical property of material for coating applications. At present, 7 - 8 wt. % Yttria incorporated into zirconia (Yttria stabilized zirconia) is commercially recognized as potential candidate for TBC material [8]. But at the elevated temperature (T >1200 °C), it exhibits a phase transformation and become less effective due to induced thermal stresses, reduction in coating porosity as well as adherence between the bond coat and YSZ layer [9,10]. Due to the limitations of operating temperature, YSZ's potential has been saturated, and current focus of researchers is on the development of new TBC material for higher operating temperatures [11]. Pyrochlore structured TBC materials with extremely good thermal and structural phase stability at high temperature are promising candidates for the applications having operating temperatures above 1300 °C. Among pyrochlore structured ceramic, some rare earth based zirconates ($Ln_2Zr_2O_7$), have received the most attention as TBC material where Ln can be La, Sm, Nd, Gd, Yb, and Eu [12–15]. Pyrochlore structured zirconates exhibit the lowest thermal conductivity among different fluorite structured ceramics. In addition, ceria [$La_2Ce_2O_7$, $La_2(Zr_{0.7}Ce_{0.3})_2O_7$] and hafnia [$Gd_2Hf_2O_7$, $La_2Hf_2O_7$] based pyrochlore were also investigated for TBC applications [16]. $La_2Zr_2O_7$ pyrochlore with low thermal conductivity (1.56 W/m-K), low sintering tendency, and thermal stability up to 2000 °C are the most promising TBCs comparable to YSZ, but it has lower thermal expansion coefficient which leads to thermal stresses. Pyrochlore $Gd_2Zr_2O_7$ has not only high thermal expansion but also low thermal conductivity can be used as TBC materials [11]. In an advanced nuclear reactor, new transuranium actinides have been generated on the neutron absorption by $^{238}U$. Therefore, inert matrix (IM) with low thermal conductivity is required for high power levels during irradiation and high concentration of actinides [17]. Pyrochlore structured ceramics are also potential candidate for inert matrix (IM) fuel applications due to excellent thermodynamic properties and high-temperature stability under ion irradiation. Various pyrochlore structured oxides have stability up to high-temperature close to 2300 °C, is selected for present study. $Gd_2Zr_2O_7$ shows the pyrochlore to defect fluorite phase transition around 1560 °C whereas, $Nd_2Zr_2O_7$ is stable up to 2300 °C [18]. Furthermore, heat transport mechanism in pyrochlore materials needs to be investigated for their use in a high-temperature environment. Generally, pyrochlore oxides are insulators, so heat transport through electronic conduction is negligible, and radiative heat transport takes place at very high temperatures. In an insulator, the heat transport is mainly due to the phonon contribution through several scattering phenomena such as phonon-phonon, defects, grain boundary scattering *etc.* depending on the different range of temperatures [19,20]. Heat transport is governed by defect and boundary scattering at low and intermediate temperatures, while at high temperatures umklapp phonon-phonon scattering dominates. Nanocrystalline (nc) materials contain a large number of grain boundaries which are considered an obstacle in heat transport and exhibits lower thermal conductivity as compared to corresponding single crystal materials [21]. There are few reported studies related to the variation of thermal conductivity with grain size for binary oxides, thermoelectric materials, and semiconductors [22–25]. But, in the case of pyrochlore, the effect of substitution and cationic mass on the thermodynamic conductivity at low-temperature range has been reported but there is no report related to effect of porosity, grain size on the thermal conductivity of pyrochlore. Therefore, better understanding of the thermal properties in pyrochlore structured ceramic at ambient temperature and the effect of several scattering mechanisms needs to be investigated.

Here, the effects of grain size and temperature on thermal conductivity of nc-$Gd_2Ti_2O_7$ pyrochlore oxide fabricated by spark plasma sintering method, is investigated with aim to understand the effects of porosity and structure on the thermal behavior of nanocrystalline pyrochlore.

**Experimental Details**

**Synthesis**

$Gd_2Ti_2O_7$ pyrochlore of varied grain size and porosity are prepared by spark plasma sintering (SPS) technique in combination with high energy ball milling (HEBM). Starting reagents $Gd_2O_3$ and $TiO_2$ having purity of 99.9% (Sigma-Aldrich) were calcinated in the air at the temperature of 800 °C for 12 hours to evaporate water impurity. According to stoichiometry (1 $Gd_2O_3$ : 2 $TiO_2$), appropriate amounts of the starting regents were weighed out and grinded with zirconia balls (d -1 mm) using HEBM system (Fritsch, Pulverisette 7, Idar-Oberstein, Germany). The milling was carried out for 80 rounds having each round of 30 minutes at the speed of 500 rpm. To avoid excessive heat, milling was also stopped for 15 minutes after each cycle, and the measured value of the temperature of the powder was about 50 °C during milling. After milling for a duration of 20, 30, and 40 hours, samples were taken out and diffraction patterns were recorded to examine the structural analysis.

Amorphous powder after high energy ball milling was integrated to a compact single phase pyrochlore nano-ceramic having varied grain size and porosity by SPS sintering (Dr. Sinter® SPS-211 Lx system) at different temperatures. To get compact powder geometry, SPS allows a high electric current to pass through the material to eliminate the inter-particle pores. For every spark plasma sintering, ~1 g nano-ceramic powder was transferred to a graphite die (dia. ~10 mm) and it was enwrapped into a graphite paper (~0.20 mm) for proper fitting and to enhance the conduction between the assembly and powder. Sintering was carried out at different temperatures of 800 °C, 1000 °C, and 1200 °C, and pressure of 40 MPa with a dwelling time of 20 minutes. A post-sintering annealing is also carried out at 100 °C lower than the sintering temperature for duration of 20 minutes to avoid the cracking of the pellets. All the samples were annealed for 12 hours in air atmosphere at 800 °C in order to detach the carbonaceous impurities attach from the graphite die.

**Characterization**

Structural analysis was take place by using a Rigaku miniflex 600 diffractometer (CuK$_α$ radiation with λ=1.5460 Å). The XRD pattern was recorded in the angle range from 10° - 90° with a scan rate of 1° per minute at an increment of 0.010°. Microstructure characterization of as-milled powder, sintered, and annealed specimens was carried out using a field emission scanning electron microscopy (FESEM - Carl Zeiss Supra 55 Jana, Germany). A Denton Sputterer (Denton Vacuum, LLC, NJ, USA) was used to deposit a thin layer of platinum in an argon atmosphere on the fractured surface to get better morphology results. Raman measurements have been carried out using WiTec alpha 300 RA system equipped with a CCD and Nd-YAG laser of 532 nm to excite the sample at the advanced instrumentation facility (AIRF), JNU New Delhi. The thermal diffusivity measurements of the sintered pellets of $Gd_2Ti_2O_7$ were performed in a relation with temperature using a Laser Flash Analyzer (LFA 427, Netzsch Instruments, Germany). A laser beam is focused on the surface of the pellet during LFA measurement and rise in heat is calculated by measuring the infrared blackbody radiation on other side of pellet. Diffusivity values were obtained through the room temperature to 900 °C upon heating. At each temperature step, three laser shots were collected and averaged to determine the values of thermal diffusivity. The values of thermal diffusivity were calculated by fitting the rise in temperature signals using a modified Cape-Lehman Model [26]. The nc-$Gd_2Ti_2O_7$ specimens were coated with graphite thin films to enhance the emissivity of both surfaces of the pellets, and absorption of laser light respectively. All experiments were carried with approximately $10^{-15}$ ppm flow of oxygen at a rate of 100 ml/min under gettered UHP argon. The thermal conductivity values were determined as follows

$$k = \rho.C_p.\lambda \ldots\ldots(1)$$

Where $\rho$, $C_p$, and $k$ are density (kg.m$^{-3}$), specific heat capacity at constant pressure (J.kg$^{-1}$.K$^{-1}$) and thermal conductivity (W.m$^{-1}$.K$^{-1}$), respectively. Moreover, $\lambda$ is defined as thermal diffusivity in (mm$^2$/s). The density at room temperature was calculated using the Archimedes principle and it was corrected using the density kit apparatus for the high temperatures.

**Results and Discussion**

XRD pattern along with SEM images of the as-milled nano-powders showed that nano-powders are amorphous in nature and micro-sized granules contain nano-sized particles. No impurity peaks confirm the homogenous mixing of all the reagents using HEBM. **Fig. 1** indicates the SEM micrographs of the fractured surfaces of nc-$Gd_2Ti_2O_7$ specimens sintered at temperatures of 800 °C, 1000 °C, and 1200 °C. At the lower sintering temperature (800 °C), large number of smaller grains present can be easily seen, so accurate calculation of the grain size is difficult.

**Table 1: Porosity and crystallite size of the nc-$Gd_2Ti_2O_7$ for the specimens as-sintered at different temperatures for the duration of 20 minutes.**

| Annealing temperature | 800 °C | 1000 °C | 1200 °C |
|---|---|---|---|
| **Porosity (%)** | 40 | 20 | 5 |
| **Micro-strain (%)** | 0.115±0.009 | 0.101±0.007 | 0.091±0.006 |
| **Crystallite size (nm)** | 48±5.81 | 64±8.26 | 84±12.20 |

The SEM images of the nc-$Gd_2Ti_2O_7$ specimen sintered at 1000 °C and 1200 °C indicate an increase in grain size and reduction in density of sub-micron-sized pores with increasing the sintering temperature. The calculated values of the crystallite size reported in **Table 1** shows the average grain growth with an increase in sintering temperature and highly dense microstructure are formed at higher sintering temperature. Formation of bigger crystallite size as compared to ball-milled powders confirms that grain growth itself occurs during sintering. It also displays that pores of sub-micron size appeared at grain boundaries at sintering temperature of 1200 °C. Temperature-dependent grain growth kinetics was reported for the polycrystalline $La_2Zr_2O_7$, annealed at different temperatures for various annealing times [27]. Our earlier studies of the nc-$Gd_2Ti_2O_7$ showed that grain growth critically depends upon the annealing temperature, time and size of the nanoparticle [28]. Present sintering studies on the amorphous nano-powder also exhibits similar grain growth indicates that grain growth occur in the both crystalline as well as amorphous ceramic.

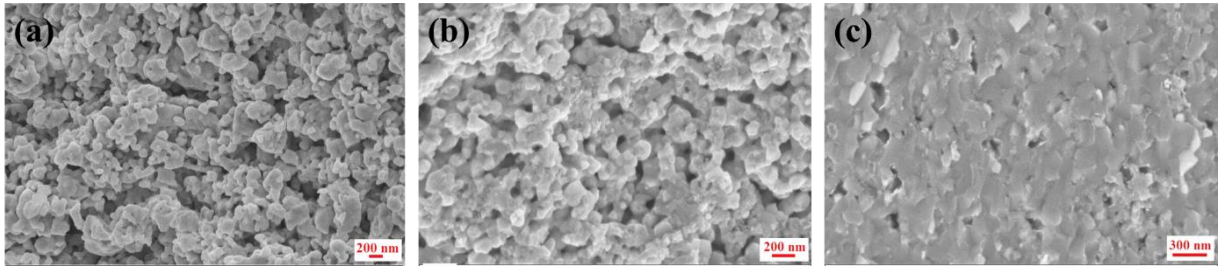

**Fig. 1.** SEM micrographs of nc-$Gd_2Ti_2O_7$ specimens sintered at temperature (a) 800 °C, (b) 1000 °C, and (c) 1200 °C confirm that highly dense grains are formed at higher sintering temperature.

**Fig. 2.** shows the refined XRD pattern of all the three samples sintered at different temperatures for the duration of 20 minutes. XRD pattern displays the superstructure peaks of pyrochlore phase having unit cell length of 10.1894 Å. These superstructure peaks of very small intensities are formed due to the extra conditions of allowed reflections (h = 2n+1, h, k, l = 4n+2 or l = 2n). It is also noted that sub-cell peaks of fluorite structure forms due to the reflection conditions of h+k, l+k, l+h are a multiple of four (k+l = 4n, h+k = 4n, l+h = 4n) and are consistent with reported results [29]. The intensity of these peaks was increased, and peak shifting towards a higher angle was observed for the high-temperature annealed samples. The reduced full width at half maxima (FWHM) of the diffraction peaks indicates the increased grain size at high sintering temperatures. The values of crystallite size and micro-strain were obtained using a linear plot between β.cos(θ) *vs* sin(θ) in the Williamson-Hall analysis [30]. The crystallite size was increased upon sintering as a result of temperature-induced grain growth. To get the peak positions and FWHM of the diffracted peaks, five peaks of higher intensities were fitted using the pseudo-Voigt function in the W-H analysis. **Fig.2(c).** shows that variation of micro strain and crystallite size with increasing

sintering temperature. The calculated crystallite size shows the thermally induced grain growth either by curvature-driven boundary displacement mechanism or grain-rotation mechanism [31]. An increase in the crystallite size with increasing temperature was also reported for the $Eu^{3+}$ doped $Gd_2Ti_2O_7$ thin films [32].

Structural analysis of all the three samples has been performed using powder x-ray diffraction and Raman spectroscopic observations. The diffraction pattern of all the three samples has been refined using Rietveld refinement program Fullprof-2005. The diffraction peaks grow with sintering temperature are showed with corresponding Miller indices (h, k, *l*). Pseudo-Voigt function and six-coefficient polynomial function were used to fit the diffraction peaks and background, respectively. XRD pattern indicates that all three synthesized crystalline samples are in single-phase having pyrochlore superstructures in **Fig. 2(a)**. In the Pyrochlore structure with composition $A_2B_2O_7$, both cations of A- and B-site lie at the Wyckoff sites 16d (1/2, 1/2, 1/2) and 16c (0, 0, 0) with equal "$\bar{3}m$" symmetry respectively [33]. The pyrochlore superstructure having two oxygen atoms, one at 48f (x, 0.125, 0.125) with two $A^{3+}$ and two $B^{4+}$ neighbors with symmetry "mm", and another at 8b (0.375, 0.375, 0.375) with four $A^{3+}$ neighbors with site symmetry "43m." The crystallographic site 8a (1/8, 1/8, 1/8) for third anion surrounded by four $B^{4+}$ cations with symmetry "$\overline{43}m$" increase relaxation to 48f anionic oxygen toward the vacant site to compensate the charge. The value of crystallographic variable "*x*" for 48f oxygen plays an important role in deciding the stability of structure and its value varies from 0.3125 to 0.3750. A-site cations are associated with six 48f plus two 8b anions, while B-site cations are linked with six 48f anions, respectively. In the case of pyrochlore superstructure, the value of A-$O_{48f}$ bond length is always found larger as compared to A-$O_{8b}$ bond length and becomes similar for an anion-deficient fluorite structure, *i.e.,* when x=0.3750. Rietveld refinement reveals that the bond length decreases on

increasing sintering temperature. This effect can be correlated by with shifting in the position of phonon

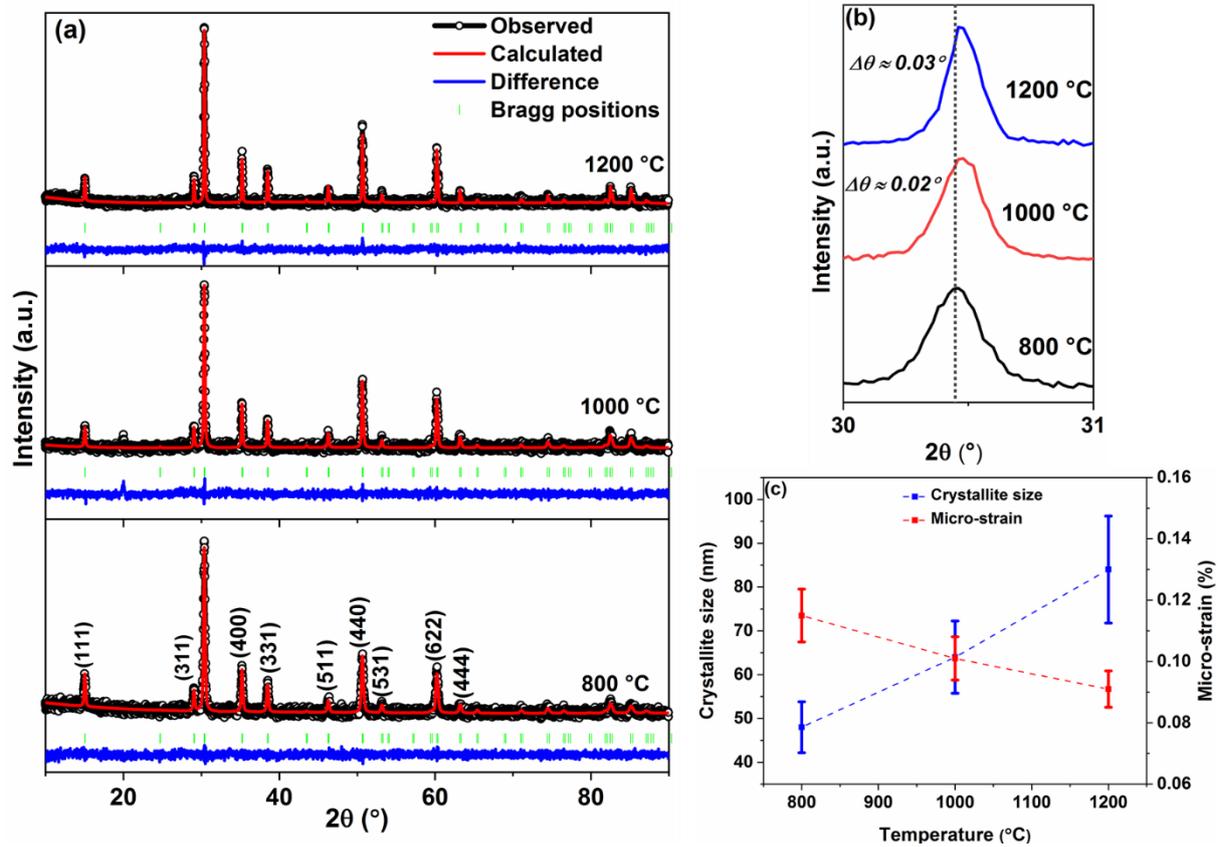

**Fig. 2.** Indicating **(a)** refined x-ray diffraction pattern, **(b)** zoomed XRD pattern of the most intense peak (222), **(c)** Variation of the crystallite size and micro strain with increasing sintering temperature for the nc-$Gd_2Ti_2O_7$.

vibration towards high wave number side as reported in **Table 3.** The increased bond length A-$O_{8b}$ reveals the information about strengthening the pyrochlore in structure or the reduction in the displacement of $O_{48f}$. All the specimen shows the different extent of pyrochlore superstructure and also observed the shifting in the diffraction peaks towards the higher angle on increasing the sintering temperature. There is angle shift of $\Delta\theta \sim 0.02°$ and $\Delta\theta \sim 0.03°$ for the specimen synthesized at 1000 °C and 1200 °C with respect to the sample 800 °C **[Fig. 2(b)]**, which implies that nc- $Gd_2Ti_2O_7$ sintered at 800 °C has the largest lattice parameter and sample sintered at 1200 °C has the smallest lattice parameter as shown in **Table 2**. For structural analysis, Raman spectra for all the samples were performed in the range of wave number 150-1000 cm$^{-1}$. The results give the details about the change in the anionic sublattice position of oxygen atom. Theoretically, six vibration modes for $Gd_2Ti_2O_7$ designated as

| Table 2. Parameters calculated from the Rietveld refinement for the nc-$Gd_2Ti_2O_7$ | | | |
|---|---|---|---|
| **Pyrochlore superstructure** | **800 °C** | **1000 °C** | **1200 °C** |
| **Lattice parameter (Å)** | 10.1894 (8) | 10.1890 (6) | 10.1855 (4) |
| **O$_{48f}$** | 0.3256 (2) | 0.3259 (3) | 0.3221 (5) |
| **Gd-O$_{8b}$ bond length (Å)** | 2.2061(1) | 2.2060 (1) | 2.2052 (1) |
| **Gd-O$_{48f}$ bond length (Å)** | 2.5274 (143) | 2.5280 (100) | 2.5545 (76) |
| **Ti-O$_{48f}$ bond length (Å)** | 1.9607 (80) | 1.9602 (56) | 1.9446 (58) |
| $\chi^2$ | 1.009 | 1.126 | 1.123 |

$E_g$, $A_{1g}$, and $4T_{2g}$ can be observed in $Gd_2Ti_2O_7$ pyrochlore. In the case of nc-$Gd_2Ti_2O_7$, only five Raman active vibration modes ($A_{1g}$, $E_g$, and $3T_{2g}$) are contributed due to presence of oxygen at 48f and 8a sites. The eminent vibration bands $E_g$ (~309 cm$^{-1}$), and $A_{1g}$ (~517 cm$^{-1}$) appears as a result of O-Gd-O bonding and Gd-O stretch [34,35]. Rest of the $T_{2g}$ modes of vibration corresponding wavenumber of ~222 cm$^{-1}$, and ~550 cm$^{-1}$ shows good agreement with the reported results [36]. At lower wavenumber, Ti-O stretching Raman active modes were not appeared due to small bond distance of Ti-O bond. The appearance of theoretically-Raman modes are in favored with the XRD results for the ordered $Gd_2Ti_2O_7$ pyrochlore structure. Raman spectrum of nc-$Gd_2Ti_2O_7$ is similar to bulk $Gd_2Ti_2O_7$ pyrochlore showing the no significant change in structural ordering and changed grain microstructure of the material is observed **[Fig.3 (a)].**

**Table 3. Experimentally measured Raman frequencies for the nc-Gd$_2$Ti$_2$O$_7$ sintered at different temperatures.**

| Pyrochlore Superstructure | T$_{2g}$ (cm$^{-1}$) | E$_g$ (cm$^{-1}$) | A$_{1g}$ (cm$^{-1}$) | T$_{2g}$ (cm$^{-1}$) |
|---|---|---|---|---|
| **800 °C** | 222 | 308.9 | 516.8 | 550.2 |
| **1000 °C** | 222.5 | 309.1 | 516.9 | 548 |
| **1200 °C** | 220.7 | 310.3 | 517 | 550.8 |

Sanjuán *et al.* has been also reported the same features for the high energy milled Gd$_2$Ti$_2$O$_7$ pyrochlore, and observed the amorphous fraction in as-milled powder [37]. The locality of Raman active modes persist almost same with changing sintering temperature upon high temperature sintering [38,39]. **Fig. 4.** shows the thermal diffusivity of nc-Gd$_2$Ti$_2$O$_7$ in the temperature range from room temperature to 900 °C. Thermal diffusivity of each sample is decreasing with increase in the temperature. This follows an exponential decay behavior and fitted with the following function (2) -

$$y_o = A.\exp\left(-\frac{x}{t}\right) + y_o \ldots\ldots (2)$$

Where A is amplitude, $y_o$ is offset and t is the time constant are the parameters arising from the fitting of thermal diffusivity values as shown in **Table 4**. Increasing offset value with temperature shows the value of thermal diffusivity for the specimen sintered at 1200 °C is much larger compared to sintered at 800 °C.

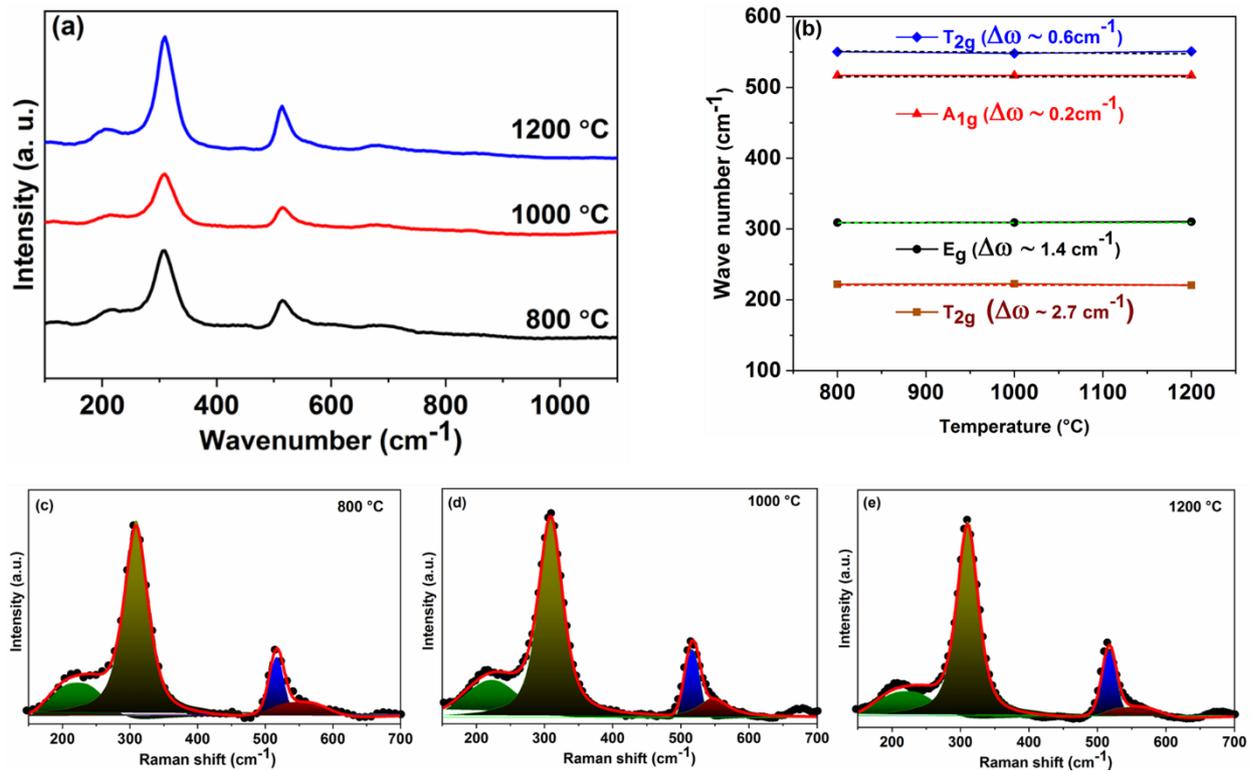

**Fig.3.** Representative (a) Micro-Raman scattering results for the for the nc-$Gd_2Ti_2O_7$, (b) Variation of the wave number corresponding to different bands for all the samples. **(c-e)** cumulative peak fitting for the samples sintered at different temperatures.

Thermal diffusivity of lower crystallite sample shows higher decay rate as compared to higher crystallite samples. On comparison, it is noted that the decay rate is similar for the specimens sintered at 1000 °C and 1200 °C due to crystallite size. At starting, thermal diffusivity shows higher value for the sample sintered 1200 °C and with increasing temperature, it shows similar decay rate with the sample sintered at 1000 °C. However, the rate of reduction in the thermal diffusivity critically depends upon the sintering temperature. The rate of decrease in the thermal diffusivity at high temperatures is almost for same grain size specimens. The reduction in the thermal diffusivity at high temperatures is due to the shortened mean free path of phonon with increasing temperatures which leads to increase in the phonon scattering resulting in low thermal diffusivity. The comparison of the observed values with the reported values of thermal diffusivity for $Gd_2Ti_2O_7$ sintered at 1600 °C, shows the similarity for the largest grain size sample sintered at 1200 °C [40]. Ibrahim *et al.* investigated the grain size-dependent thermal diffusivity for the strontium titanate and revealed an enhancement in the thermal diffusivity with the increase grain size due to reduction of the grain boundaries which is consistent with data of present study.

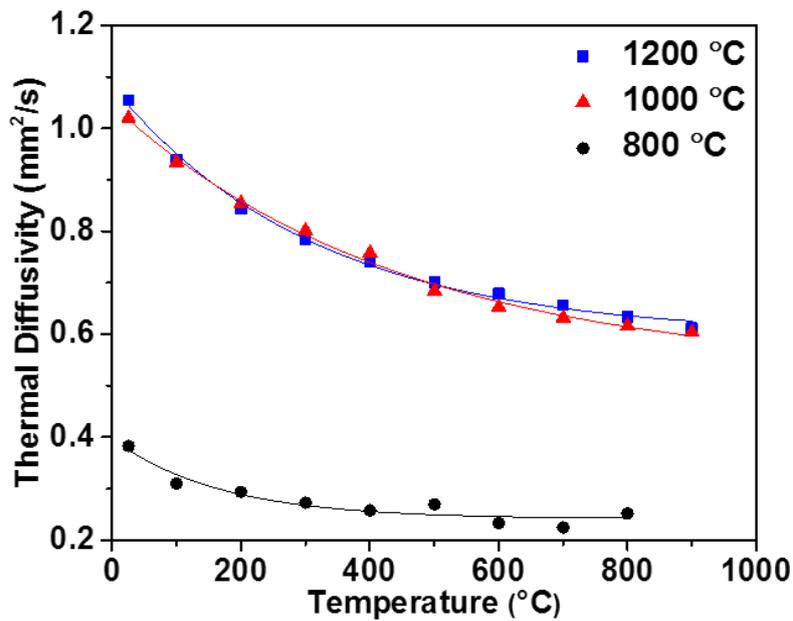

**Fig.4.** Temperature dependence on thermal diffusivity of nc-$Gd_2Ti_2O_7$ having different crystallite sizes. Thermal diffusivity is systematic decreased with increasing temperature.

The thermal diffusivity displays a systematic reduction with the increasing sintered temperature. The observation of the lower thermal diffusivity for samples annealed at lower sintering temperature can be explained on the basis of phonon scattering. The phonon scattering for smaller grain size is very low due to low density of grain boundaries and defects and imperfections are dominate [41]. Further, lower grain size sample is porous in nature and the porosity also significantly reduce the thermal diffusivity by reducing thermal conduction pathways [42]. **Fig.4.** also displays that thermal diffusivity for highly dense samples sintered at 1200 °C have the much larger values than sample sintered at 800 °C. In addition, thermal diffusivity behavior for the nc-$Gd_2Ti_2O_7$ sintered at 1000 °C and 1200 °C is similar due to the small differences in their crystallite size values. The lower densities of nc-material are responsible for the reduction of thermal diffusivity compared to single-crystal (no grain boundary scattering) materials, especially at high temperatures. Smaller nano-sized grain leads predominantly to phonon-boundary scattering. Ibrahim *et.. al.* also observed that thermal diffusivity decreases with the increasing porosity, due to increasing phonon-defects scattering [41]. With increasing grain size, the thermal diffusivity increases due to decreasing grain boundary scattering.

**Table. 4. Fitted parameters of thermal diffusivity and conductivity for the nc-$Gd_2Ti_2O_7$**

|  | Thermal Diffusivity | | | Thermal conductivity | |
| --- | --- | --- | --- | --- | --- |
|  | $y_o$ | A | t | Slope | Intercept |
| **800 °C** | 0.241±0.010 | 0.155±0.019 | 168.88±53.36 | 0.577±0.019 | -1.28E-4±4.08E-5 |
| **1000 °C** | 0.526±0.023 | 0.518±0.019 | 449.98±48.50 | 1.671±0.020 | -5.09E-4±3.91E-5 |
| **1200 °C** | 0.597±0.012 | 0.484±0.012 | 316.09±25.03 | 1.673±0.014 | -4.76E-4±2.72E-5 |

The variation in the thermal conductivity with temperature for samples having different grain is shown in **Fig.5**. It follows the linearly decreasing behavior with increasing temperature and the graphs are fitted with straight line. It indicates the reduced thermal conductivity with rising temperature and an increase in thermal conductivities with increasing grain growth in the nc-$Gd_2Ti_2O_7$. The calculated thermal conductivity value for the lower grain sized sample (sintered at 800 °C) is significantly lower as compared to the higher grain sized sample (sintered at 1000 °C). Although the thermal conductivity calculated at 1000 °C and 1200 °C is almost equal because both samples have the approximately same porosity or grain size as observed from SEM micrographs. In the present work, the observed values of thermal conductivity are also matched with minimum thermal conductivity value calculated by DFT framework based LDA and GGA methods for $Gd_2Ti_2O_7$ [43]. For insulator oxides, phonon conduction majorly plays an important role in heat transport at lower temperatures, and phonon scattering is gradually enhanced with increasing temperature. The crystal structure is fully occupied at sufficiently high temperatures, and more phonon-phonon interaction resulting the shortened mean free path of the phonons. For nanocrystalline materials, the change in the values of thermal conductivity with grain sizes can be explained mainly by boundary scattering phenomena. There are various possible layouts apart from grain boundary scattering in the nc-materials as compared to bulk materials. Nanocrystalline materials contains a large volume of grain boundaries which behaves as a barrier in the heat transmission mechanism. These grain boundaries behave as a scattering centers for phonons, that lead to decreasing thermal conductivity [20]. According to the reported results, the thermal conductivity of nanocrystalline ceramics is much lower as compared to the corresponding single crystal material [44].

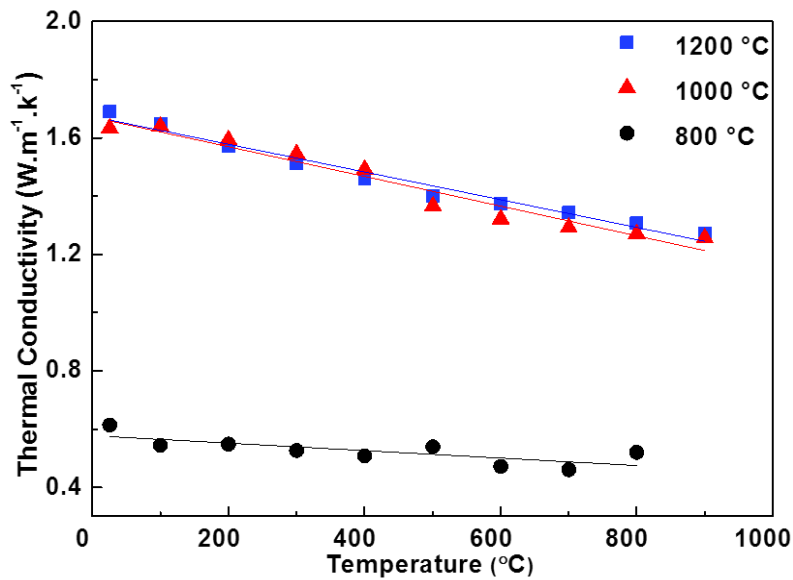

**Fig.5.** Thermal conductivity for the nc-$Gd_2Ti_2O_7$ specimens sintered at different temperatures indicates that conductivity decreases with lowering crystallite size.

At high temperatures, the increased boundary scattering resulting decrease in the thermal conductivity. The grain size dependent thermal conductivity for nc-pyrochlore structured oxides has not been studied yet experimentally. Some reported results show the variation of thermal conductivity with grain size for binary oxides like $UO_2$ [22]. Shrestha *et al.* investigated the dependence of grain size on the thermal conductivity for $UO_2$ which shows that boundary scattering is the primary source of thermal resistance. They observed that grain boundary scattering has a major impact on thermal conductivity of nc-oxides and the porosity has a large impact in case of polycrystalline samples. The presence of high porosity in $UO_2$ in polycrystalline sample has reduced the thermal conductivity, especially at higher temperature [25,45]. Liu *et al.* observed a significant reduction in the value of thermal conductivity for nanocrystalline materials with grain size and boundary scattering effect found to be more dominant with the lowering grain size [46]. In addition to grain boundary scattering phenomena, they also reported comparatively smaller contribution of the other factors like defects and porosity in reducing thermal conductivity, especially at intermediate temperatures. The grain boundary scattering dominates when the grain size is comparable to the phonons mean free path [47]. In general, pores plays salient role in phonon scattering resulting to reducing the thermal conductivity by decreasing the phonon mean free path. In nc-materials, phonon scattering is mainly caused by grain boundary scattering, phonon-pores scattering, and phonon-vacancies scattering. For high porosity samples, the pores can lead to distinct phonon scattering, which can seriously interrupt the transport of phonons in the materials [48]. According to molecular dynamics simulation results, porosity reduces the thermal conductivity by decreasing phonon conduction channels and increasing phonon scattering at the surface of pores [49,50]. In present study, lower value of thermal conductivity of sample sintered at lower temperature containing high porosity in nc-$Gd_2Ti_2O_7$ is observed and that increases with the decreasing porosity. With the increasing temperature, the density of the specimens increases resulting in higher thermal conductivity values. The sample sintered at 1200 °C containing the lowest porosity shows a high thermal conductivity value compared to the sample containing high porosity. Based on reported results for nc-materials, we expect that phonon-pores scattering plays a significant role in the nc-$Gd_2Ti_2O_7$ at intermediate temperatures up to 500 °C. The significant dispersal of grain size and ineludible appearance of porosity in nc-$Gd_2Ti_2O_7$ result to reduced thermal conductivity by increasing phonon scattering. At high temperatures, boundary scattering dominates with increasing temperature.

**Conclusions**

In summary, nc-$Gd_2Ti_2O_7$ have synthesized by spark plasma sintering combined with high-energy ball milling and characterized by X-ray powder diffraction (XRD), Raman spectroscopy, and scanning electron microscope (SEM) techniques. The grains increase gradually with the temperature. A systematic investigation of the grain-size dependent thermal properties of nc-$Gd_2Ti_2O_7$ showed besides boundary scattering, defect scattering also plays a significant role in nc-materials. It has been seen that thermal conductivity at the starting temperature is reduced due to phonon-pore scattering while boundary scattering dominates at high temperatures. The detailed study of the scattering mechanism in nc-$Gd_2Ti_2O_7$ would be helpful for the researchers working on modeling thermal insulation materials. Thus, understanding the dependence of grain size on thermal conductivity would also not only be beneficial for the researchers working on modeling thermal insulation materials but also to studying the heat transmission mechanism in applied materials, especially nc-materials.

**Acknowledgment**

One of the authors (P.K.K.) acknowledges USIEF for awarding postdoctoral research fellowship and Board of Research in Nuclear Sciences (BRNS) for providing financial support under the project 58/14/05/2019-BRNS/37013. Yogendar Singh acknowledges the Prime Minister's Research Fellowship for awarding fellowship with reference number PMRF -2122-2836. The authors also acknowledge the Central Instrumentation Facility (CIF), School of Physical Sciences (SPS) for XRD measurements.